\documentclass[aps,prd,onecolumn,showpacs,showkeys,12pt]{revtex4-1}

\usepackage{graphicx,color}
\usepackage{rotating}
\usepackage{amssymb}
\usepackage{amsfonts}
\usepackage{amsmath}
\newcommand{\be}{\begin{equation}}
\newcommand{\ee}{\end{equation}}
\newcommand{\bea}{\begin{eqnarray}}
\newcommand{\eea}{\end{eqnarray}}

\begin{document}


\title{Minimal length and the flow of entropy from black holes\footnote{Awarded ``Honorable Mention" in the 2018 Gravity Research Foundation Essay Competition��} }

\author{Ana Alonso-Serrano}
\email[]{ana.alonso.serrano@aei.mpg.de}
\affiliation{Max Planck Institute for Gravitational Physics, Albert Einstein Institute, Am M\"{u}hlenberg 1, D-14476 Golm, Germany}

\author{Mariusz P. D\c{a}browski}
\email[]{mariusz.dabrowski@usz.edu.pl}
\affiliation{Institute of Physics, University of Szczecin, Wielkopolska 15, 70-451 Szczecin, Poland}
\affiliation{National Centre for Nuclear Research, Andrzeja So{\l}tana 7, 05-400 Otwock, Poland}
\affiliation{Copernicus Center for Interdisciplinary Studies, S{\l}�awkowska 17, 31-016 Krak\'ow, Poland}

\author{Hussain Gohar}
\email[]{hussain.gohar@usz.edu.pl}
\affiliation{Institute of Physics, University of Szczecin, Wielkopolska 15, 70-451 Szczecin, Poland}
\date{\today}
\begin{abstract}
The existence of a minimal length, predicted by different theories of quantum gravity, can be phenomenologically described in terms of a generalized uncertainty principle. We consider the impact of this quantum gravity motivated effect onto the information budget of a black hole and the sparsity of Hawking radiation during the black hole evaporation process. We show that the information is not transmitted at the same rate during the final stages of the evaporation, and that the Hawking radiation is not sparse anymore when the black hole approaches the Planck mass. 

\end{abstract}


\maketitle


The appearance of black hole thermodynamics, as a result of the analogy between thermodynamics and black hole mechanics, have provided very interesting tools to develop completely new directions of research in the last half century. The possibility to define a notion of temperature associated with the black hole (Hawking temperature \cite{hr1a}) and a corresponding entropy (Bekenstein entropy \cite{bentropy}) - both being related to an area of the event horizon - reflects the fact of considering the black hole as an emitter which can evaporate. This mechanism of evaporation relies in the very realm of quantum field theory, and, may be, one of its more interesting features is that it seems to imply a non unitary evolution, which gave rise to the well known problem of information loss paradox, that remains being a mystery \cite{hr1b,hr22,hr3,hr11a,hr22b,epr}
	
It is interesting to note here that the entropy can be understood as lack of information about the internal configuration of the system \cite{bentropy,shannon}, measuring the inaccessibility of information to an external observer. This is directly a reflection of the emergent macroscopic properties arising from the quantum statistical mechanics underlying the behavior of the quantum microstates. In order to completely understand the origin of this entropy and the nature of these microstates, it would be necessary to have a complete theory of quantum gravity that it is still an elusive theory.
	
Considering the case of having a Schwarzschild black hole with mass $M$, the Hawking temperature and Bekenstein entropy are expressed, respectively, as
\be
T=\frac{c^2}{8\pi k_B}\frac{m^2_p}{M}, ~~~~ S=4\pi k_B\left(\frac{M}{m_p}\right)^2=k_B \hat S~\label{BH}
\ee
where $k_B$ is the Boltzmann constant, $m_p$ is the Planck mass, $c$ is the speed of light, and $\hat{S}$ is the entropy measured in nats.

We are interested in understanding the entropy flux in the Hawking radiation in order to deal with information issues related to black hole evaporation. As a first step in the study of the entropy flux in the radiation, it has been shown \cite{ana1} that a black-body emits a budget of average entropy of  $3.9 ~\text{bits/photon}$, that, as it is a unitary process, it is compensated by the same amount of hidden information in correlations. Interestingly, for a Schwarzschild black hole, the loss of Bekenstein entropy is exactly equal to the information content emitted from a black body and is given by \cite{ana2}
\be
		\frac{d\hat S_2}{dN}\approx 3.9 ~\text{bits/photon},
		\label{lump}
		\ee
where $\hat S_2=\hat S/\text{ln2}$, and $N$ is the total number of particles emitted from a black hole, so the Hawking radiation is analogous to the black-body radiation. It has also been shown that the Bekenstein entropy of a black hole is completely transfered to a gain of Clausius entropy of the Hawking radiation field, giving a complete coherent picture of entropy fluxes. In addition, it is interesting to note that the total number of emitted quanta $N$ is given only in terms of the initial mass (or analogous, the initial Bekenstein entropy):
\begin{equation}
		N= \frac{30\xi(3)}{\pi^4}\left(\frac{M}{m_p}\right)^24\pi k_B\approx 0.26 \; \hat S_2.
		\label{non-GUP}
\end{equation}

Another important feature to understand the Hawking flux, is that it is very sparse \cite{matt1}, which means that the average time between emission of Hawking quanta is very large compared to the timescale of the energies of these quanta. The sparsity of Hawking radiation continues throughout the evaporation process from its early stages to the late stages. It can be defined by some different dimensionless parameters $\eta$, which are suitable for the comparison of the mean time between emission of successive Hawking quanta with several natural timescales that can be associated  with the emitted quanta. A general parameter is defined by \cite{matt1}
\begin{equation}
\eta =C\frac{\lambda_{\text{thermal}}^2}{gA},
\end{equation}
where $C$ is a dimensionless constant that depends on the specific parameter ($\eta$) we are choosing \cite{matt1}, $g$ is the spin degeneracy factor, $A$ is the area, and $\lambda_{\text{thermal}}=2\pi\hbar c/(k_BT)$ is the ``thermal wavelength''. For a Schwarzschild black hole, the temperature in the thermal wavelength is given by the Hawking temperature and the area should be replaced by an effective area given by $A_{\text{eff}}=(27/4) A$ \cite{matt1}. In this case, the relevant factor in any dimensionless parameter does not depend on $M$ and results in a number much greater than one
\be
\frac{\lambda^2_{thermal}}{A_{eff}}=\frac{64\pi^3}{27}\sim 73.5...\gg  1. 
\label{sparse}
\ee
Usually, for the emitters in the laboratory, the physical size of the emitter is greater than the thermal wavelength, so the dimensionless parameter is less than one, but interestingly, for black holes the behavior is completely opposite.

This provides a complete picture of the semiclassical behavior of Hawking flux. What we are interested to study here is how effects coming from the underlying theory of quantum gravity can change this picture at the last stages of evaporation, when the black hole reaches the Planck length. In order to describe these effects, we model them by the phenomenological general uncertainty principle that is model independent. The principle reflects a modification of the canonical commutator due to the existence of a minimal length when gravity is introduced into the play \cite{garay}. It appears as a prediction of different theories of quantum gravity, as can be string theory \cite{stringa,stringb,stringc}  or loop quantum gravity \cite{lqg1,lqg2}.

The generalization of Heisenberg uncertainty principle by including gravity takes the form~\cite{alpha2,GUP2,GUP3}
\be
\Delta x \Delta p = \hbar \left[1 + \alpha^2 (\Delta p )^2 \right],
\label{GUPb}
\ee
where $x$ is the position, $p$ the momentum, $l_{pl}$ the Planck length, $\hbar$ the Planck constant, and $\alpha = \alpha_0 \frac{l_{pl}}{\hbar}$ with $\alpha_0$ a dimensionless constant that describes the scale of quantum gravity effects, and thus, it is expected to be order unity.

The modification of the canonical commutator leads to a modification of Hawking temperature, when we express the momentum in terms of the minimum uncertainty in position and expand in series (at first order approximation), resulting in the expression \cite{gohar,alpha2}
\begin{equation}
T_{GUP} = T \left[ 1 + \frac{4\alpha^2 \pi^2  k_B^2}{c^2} T^2  + 2\left(\frac{4 \alpha^2\pi^2  k_B^2}{c^2} \right)^2 T^4 + \ldots \right],
\label{TGUP}
\end{equation}
and the corresponding modified Bekenstein entropy
\begin{equation}
S_{GUP}= S - \frac{\alpha^2c^2 m_p^2 k_B   \pi}{4}\ln{\frac{S}{S_0}} + \frac{\alpha^4c^4 m_p^4k_B^2   \pi^2}{4} \frac{1}{S} + \ldots,
\label{SGUP}
\end{equation}
were $S_0$, and $A_0$ are integration constants, and $T$ and $S$ are the standard Hawking temperature and Bekenstein entropy, respectively. 

It is important to note that in the consideration of parameter $\alpha$ we have taken into account a microcanonical correction, that keeps the area fixed and results in a correction over counting microstates. It would be also possible to consider a canonical correction, due to thermal fluctuations of the area, although it can be considered less fundamental in this approach \cite{collide}. On the other hand, another important remark is the generalized uncertainty principle prevents the total evaporation of black holes and predicts the existence of a final remnant, when the evaporation process ends, due to the appearance of a minimal length in the theory \cite{alpha2}.

\begin{figure}
\includegraphics[width=8.5cm]{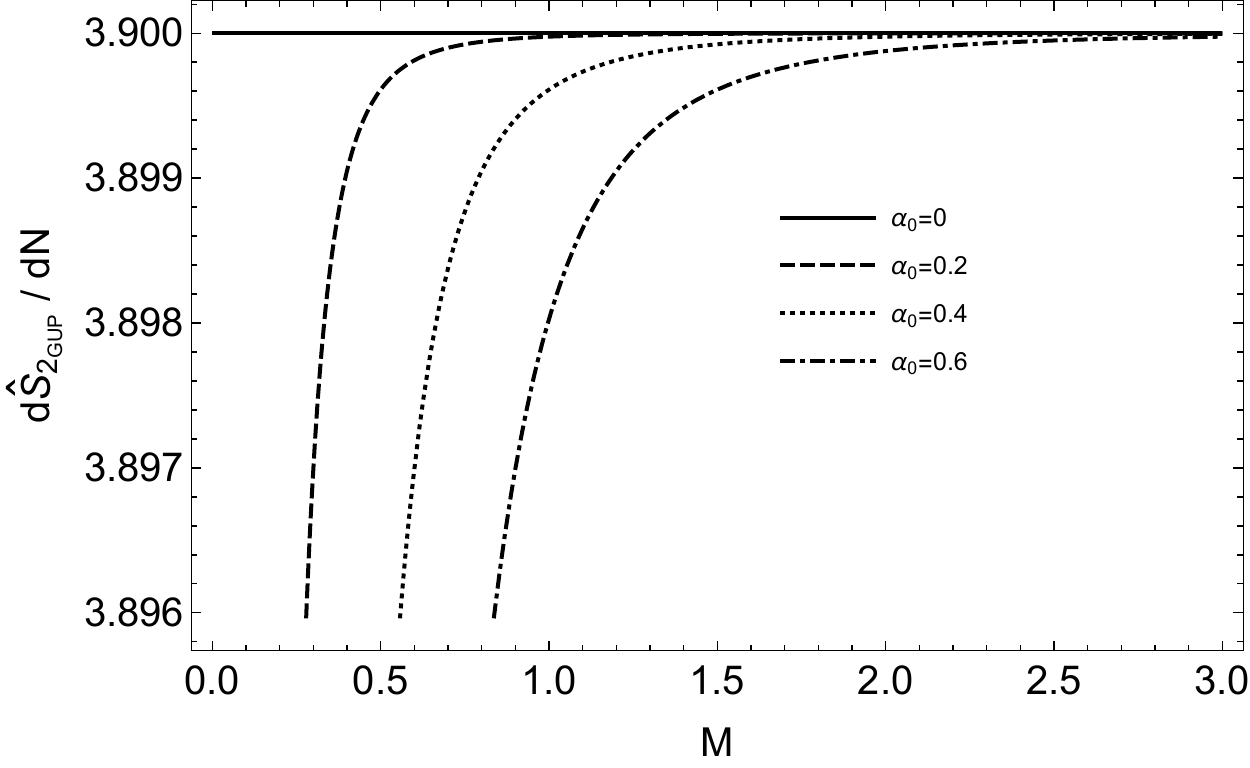}~~~
\caption{Modified Bekenstein entropy loss, $d\hat S_{2_{GUP}}/dN$ as a function of $M$ for different values of $\alpha_0$.} 
\label{fig1}
\end{figure}
From previous expressions it is possible to calculate how the generalized uncertainty principle (\ref{GUPb}) modifies the flux of average entropy per photon emitted from the black hole \cite{gohar}, as expressed by 
\be
\frac{d\hat S_{2_{GUP}}}{dN}= \frac{\pi^4}{30\xi(3)\text{ln2}}\left[1-\left(\frac{\alpha c}{4}\right)^4\left(\frac{m_p^2}{M}\right)^4+...\right]~\text{bits/photon}.
\label{dSGUPdN}
\ee
It is easy to see that the average entropy budget per particle decreases when a black hole approaches the Planck mass for different values of $\alpha_0$  (cf. Fig. \ref{fig1}). This result shows that the flow of entropy is not constant along the whole evaporation process, having less hidden information at final stages of the evaporation process. Consequently, also the total number of particles emitted from a black hole is modified to 
\be
N_{GUP}= \frac{30\xi(3)}{\pi^4}\left[\frac{4\pi}{m_p^2} M^2  -\frac{\alpha^2 c^2m^2_p\pi}{4}\text{ln}{\left( \frac{M^2}{M_0^2}\right)}\right], 
\label{NM}
\ee
where $M$ is the initial mass of a black hole and $M_0$ is an integration constant \cite{gohar}. This expression reveals that the total number of emitted particles is lower in this case than when we do not consider quantum gravity motivated effects, what is consistent with the appearance of a remnant as a final stage of the evaporation process.
\begin{figure} 
	\includegraphics[width=8.5cm]{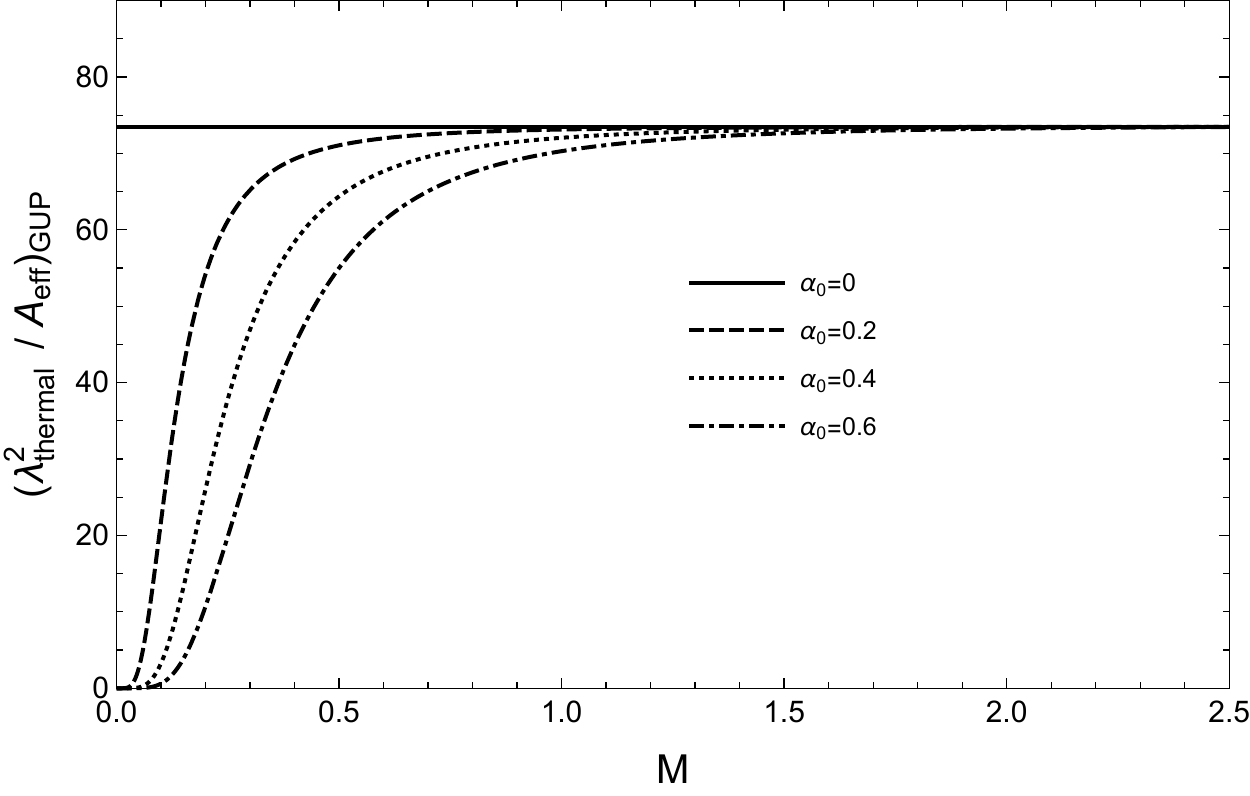}
	\caption{Modified sparsity of the Hawking flux, $(\lambda^2_{thermal}/A_{eff})|_{GUP}$ versus $M$ for different values of $\alpha_0$.} 
	\label{fig4}
\end{figure}
Finally, it is also possible to calculate how the sparsity of the Hawking flux is modified, resulting in a change of the characteristic factor into 
\bea
 \frac{\lambda^2_{thermal}}{A_{eff}}|_{GUP}=\frac{64\pi^3}{27} \left(
 \frac{M^6}{  \left[M^2-(\frac{\alpha c}{4})^2m_p^4 \ln \left(\frac{ M^2}{M^2_0}\right) \right] \left[M^{2}+(\frac{\alpha c}{4})^2m_p^4\right]^{2} }\right), 
\label{spars}
\eea
which now depends on the mass $M$ of a black hole, and on the parameter $\alpha$. This result provides a relevant effect as compared to the semi-classical case, since now it is no longer much greater than one when the black hole approaches the final stages of the evaporation process (cf. Fig. \ref{fig4}).

In summary, we have considered the impact of the minimal length (predicted from several theories of quantum gravity) as expressed in terms of a generalized uncertainty principle, onto the black hole information budget and the sparsity of the Hawking radiation. We have obtained completely new results in the regime where a black hole mass approaches the Planck mass. Our analysis seems to indicate that before the end of the evaporation process, the emitted flux from the black hole contains more information and also the flux is getting thicker before it stops being emitted. The relevance of these results are related with the black hole information paradox and how much the underlying quantum gravity theory can change the semiclassical picture.

\section*{Acknowledgements}

A. A-S. is funded by the Alexander von Humboldt Foundation. A. A-S work is also supported by the Project. No. \mbox{MINECO FIS2017-86497-C2-2-P} from Spain. The work of M.P.D. and H.G. was financed by the Polish National Science Center Grants DEC-2012/06/A/ST2/00395 and UMO-2016/20/T/ST2/00490.

\end{document}